\documentclass[journal]{IEEEtran}

\usepackage{url,hyperref,lineno,microtype,subcaption}
\usepackage{multirow}
\usepackage{array}
\usepackage{makecell}
\usepackage{siunitx}
\usepackage{graphicx}
\usepackage{cite}
\usepackage{dblfloatfix}

\begin{document}

\title{Tactile Weight Rendering: A Review for Researchers and Developers} 
\author{Rubén Martín-Rodríguez$^{{1}}$, Alexandre L. Ratschat$^{{1, 2}}$, Laura Marchal-Crespo$^{1, 2}$, and Yasemin Vardar$^{1},~\IEEEmembership{Member,~IEEE}$
\thanks{$^{1}$All authors are with the Department of Cognitive Robotics, Delft University of Technology, Delft, The Netherlands}%
\thanks{$^{2}$A.L. Ratschat and L. Marchal-Crespo are also with the Department of Rehabilitation Medicine, Erasmus MC, University Medical Center Rotterdam, Rotterdam, The Netherlands}
}

\maketitle

\begin{abstract}
Haptic rendering of weight plays an essential role in naturalistic object interaction in virtual environments. While kinesthetic devices have traditionally been used for this aim by applying forces on the limbs, tactile interfaces acting on the skin have recently offered potential solutions to enhance or substitute kinesthetic ones. Here, we aim to provide an in-depth overview and comparison of existing tactile weight rendering approaches. We categorized these approaches based on their type of stimulation into asymmetric vibration and skin stretch, further divided according to the working mechanism of the devices. Then, we compared these approaches using various criteria, including physical, mechanical, and perceptual characteristics of the reported devices and their potential applications. We found that asymmetric vibration devices have the smallest form factor, while skin stretch devices relying on the motion of flat surfaces, belts, or tactors present numerous mechanical and perceptual advantages for scenarios requiring more accurate weight rendering. Finally, we discussed the selection of the proposed categorization of devices and their application scopes, together with the limitations and opportunities for future research. We hope this study guides the development and use of tactile interfaces to achieve a more naturalistic object interaction and manipulation in virtual environments.
\end{abstract}

\section{Introduction}

Over the last decade, the usage of haptic interfaces has gained considerable attention in various applications, such as teleoperation~\cite{wildenbeest_impact_2013}, virtual reality (VR) training~\cite{gal_preliminary_2011, ozen_promoting_2021}, and neurorehabilitation~\cite{ozen_towards_2022}. Haptic interfaces are mechatronic devices that can modulate physical interaction between a human and their surroundings by displaying kinesthetic cues, i.e., information on the position of and the forces acting on a limb, and tactile cues relating to sensory information from the receptors of the skin~\cite{kuchenbecker_haptics_2018}. These interfaces can guide or restrain the user's movements and render the physical properties of an object, such as friction, temperature, stiffness, roughness, and weight~\cite{klatzky2013haptic}. Among these physical properties, weight is particularly relevant because it mediates the initial phases of object interaction and the forces applied for grasping and lifting objects~\cite{jenmalm_lighter_2006, flanagan_control_2006}. Notably, the addition of weight rendering has been shown to improve the interaction with virtual objects, e.g., improving the task performance during VR assembly tasks~\cite{carlson_evaluation_2016} and in teleoperation scenarios~\cite{wildenbeest_impact_2013}. Rendering the weight of tangible virtual objects has benefits beyond enhancing performance, e.g., weight haptic rendering is associated with enhanced motor learning of tasks involving objects with complex dynamics~\cite{danion_role_2012, ozen_towards_2022}. It has also been shown to have a positive effect on the sense of embodiment and ownership in VR~\cite{kalus_pumpvr_2023}, which in turn are associated with better performance~\cite{berg2023embodiment, odermatt2021congruency}.

Initially, the weight of virtual objects was rendered via kinesthetic haptic devices by applying forces to the limbs or fingers using grounded mechanisms~\cite{giachritsis_contribution_2010, giachritsis_unimanual_2010}. However, some of these studies pointed out reduced sensitivity compared to actual weights~\cite{giachritsis_contribution_2010} and that simulated weights were perceived as ``\textit{too artificial}'' by the users~\cite{gunther_pneumact_2019}. In addition to specific device limitations, the absence of tactile feedback, known to contribute to the perception of weight~\cite{brodie_sensorimotor_1984, jones_contribution_2006}, might be behind these limitations, as suggested by~\cite{giachritsis_contribution_2010}. Psychophysical studies have shown that the provision of kinesthetic information results in less accurate detection of small weights compared to tactile information, as well as lower discrimination between masses up to approximately 200\,g~\cite{van_beek_static_2021, minamizawa_simplified_2010}. Furthermore, it has been shown that the combination of both sensory sources yields better discrimination and detection accuracy than isolated stimuli~\cite{matsui_relative_2014, van_beek_static_2021, giachritsis_contribution_2010}. These results highlight the importance of tactile stimulation in weight perception and the need to provide multisensory haptic information to achieve accurate and compelling weight rendering.   

Researchers have explored tactile displays for weight rendering to achieve such a multisensory stimulation or provide an alternative to kinesthetic haptic devices. These devices can display tactile stimulation to simulate weights using a variety of approaches. For example, numerous works used asymmetric vibrations through vibration motors to induce a pulling sensation that can modulate the perceived weight of an object~\cite{amemiya_asymmetric_2008, choi_grabity_2017}. Another approach is to reproduce the natural skin stretch of the fingerpad upon lifting an object~\cite{minamizawa_gravity_2007, schorr_fingertip_2017}.

A recent review by~\cite{lim_systematic_2021} provides a broad overview of weight rendering approaches and associated limitations. In this review, we aim to build upon their work by presenting a deeper analysis of the weight rendering approaches through tactile stimulation and comparing relevant tactile interfaces to allow researchers and developers to perform an informed selection of the best approach for their needs. To do so, we reviewed studies on human weight perception and the approaches employed to render weight through the tactile sense. We complemented the findings of these studies with other relevant articles using the same approaches for related applications---e.g., object manipulation~\cite{schorr_fingertip_2017} and mass perception~\cite{suchoski_scaling_2018}---to gain a better understanding of the capabilities of each approach. Importantly, with the gathered information, we propose a categorization to compare the available tactile interfaces for weight rendering. We considered various criteria, including the approaches' physical properties (i.e., size and mass),  mechanical characteristics (i.e., degrees of freedom, workspace, and maximum rendering force), and perceptual features (i.e., weight and direction discrimination threshold). We also noted potential applications, determined from insights and recommendations from the literature. Then, we discussed our findings in the scope of two research questions: \emph{A) Which approaches have been used to render weight through tactile stimulation?}; and B) \emph{What are the main advantages and disadvantages of each approach?}

We hope this review can guide the use of tactile interfaces in diverse domains to achieve more accurate and coherent weight rendering. Doing so could ultimately translate into a more effective integration of tactile stimulation in haptic solutions, potentially improving task performance in VR training and teleoperation, enhancing motor learning in robot-assisted rehabilitation, and providing a more naturalistic interaction with the Virtual Environment (VE). 

The remaining part of the article is organized as follows. Section~\ref{sec:background} presents the background knowledge of the sensory mechanisms underlying weight perception, emphasizing the role of tactile information. In Section~\ref{sec:tactile_solutions}, we elaborate on existing approaches for rendering the weight of objects through tactile stimulation. In Section~\ref{sec:comparison}, we present the results from the comparison across those approaches based on the physical, mechanical, and perceptual characteristics of the tactile interfaces utilizing them. We then discuss the review's findings and the comparison, together with possible opportunities and directions for future research in the field, in Section~\ref{sec:discussion}. Finally, the outcomes and implications of our study are summarized in Section \ref{sec:conclusion}.

\section{Background} \label{sec:background}
This section presents background knowledge on measurement methods in weight perception studies, the mechanisms for weight perception, and the role of cutaneous information on human weight perception.

\subsection{Measurement of perception}\label{sec:measurement}

The field of psychophysics governs understanding of the relationship between physical attributes of stimuli and their corresponding perception. Two key concepts relevant to this review are the measurement of \textit{absolute threshold} (Detection Threshold, DT) and \textit{difference threshold} (Differenz Limen or DL). The absolute threshold represents the minimal stimulus intensity required for human perception, while the difference threshold signifies the amount of change in a stimulus needed for a \textit{Just Noticeable Difference} (JND)~\cite{jones_application_2013, gescheider_psychophysics_1997}. Notably, the \textit{Weber Fraction} (WF), which denotes the proportionality of stimulus change to its initial magnitude, is another essential term. This fraction is defined as $c=\Delta\phi/\phi$, where $\Delta\phi$ is the change in magnitude and $\phi$ is the starting magnitude of the stimulus~\cite{weber_sense_1948}. Although this value is considered constant, it has been observed to drastically increase as the magnitude of the stimulus gets closer to the absolute threshold~\cite{gescheider_psychophysics_1997}. Another metric worth mentioning is the \textit{Point of Subjective Equality} (PSE), used to indicate the point at which the magnitude of a stimulus in a specific condition is perceived to be of equal intensity as that of a reference condition.

Numerous experimental methods can be found in the literature to measure these thresholds, such as the method of constant stimuli, the method of adjustments, or the method of limits. Multiple variations arise from them, using adaptive procedures and other paradigms based on statistics and signal detection theory. In-depth review and explanation of psychophysical methods are provided in~\cite{gescheider_psychophysics_1997, jones_application_2013}.

\begin{figure}[b!]
    \centering
    \includegraphics[width=8cm]{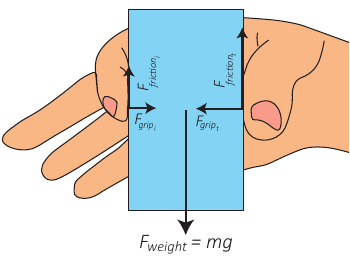}
    \caption{Schematic representation of holding an object in the air through precision grasp. Here, the object is stabilized between the thumb and index finger. The gravity, $g$, acting on the object mass, $m$, creates a downward force, $F_{weight}$. This force is stabilized by friction forces on the thumb, $F_{friction_t}$, and index fingers, $F_{friction_i}$, controlled by corresponding grip forces, $F_{grip_t}$ and $F_{grip_i}$. %The object's weight is perceived by sensory information received from the tactile receptors within the contacted skin and the proprioceptors within the muscle spindles and tendon organs on the hand and arm. 
    }
    \label{fig:weight}
\end{figure}

\newcolumntype{C}[1]{>{\centering\let\newline\\\arraybackslash\hspace{0pt}}m{#1}}

\begin{table*}[ht!]
    \centering
    \renewcommand{\arraystretch}{1.2}
    \caption{Summary of results from studies that evaluated the effect of physical factors on the perceived heaviness (i.e., weight).}\label{tab:perception}%

\begin{tabular}{cccl}
\hline
\multicolumn{2}{c}{\textbf{Factor}} &
  \textbf{\begin{tabular}[c]{@{}c@{}}Responsible Sensory System\end{tabular}} &
  \multicolumn{1}{c}{\textbf{Effect on Perceived Heaviness}} \\ \hline
  
\multirow{4}{*}{\shortstack{Biomechanical \\ conditions}} &
  Muscle fatigue &
  Kinesthetic &
    \begin{tabular}[t]{@{}l@{}}When fatigued, perceived heaviness increased due to over-estimation\\ of contraction forces and increased effort~\cite{jones_perceptions_1997}.\end{tabular} \\
 &
  Flexor sensitivity &
  Kinesthetic &
  \begin{tabular}[t]{@{}l@{}} Lifted objects were felt heavier when the sensory nerves were\\ anesthetized~\cite{gandevia_alterations_1980}.\end{tabular} \\
 &
  \begin{tabular}[t]{@{}c@{}}Grasp configuration \& style \end{tabular} &
  Kinesthetic, tactile & 
  \begin{tabular}[t]{@{}l@{}} Objects were perceived as heavier when lifted with two fingers vs. \\five, with a narrow grip vs. a wide grip, and with a small vs. large\\ contact area~\cite{flanagan_coming_2000}.\end{tabular} \\&
    \begin{tabular}[t]{@{}c@{}}Lifting method\end{tabular} &
  Kinesthetic, tactile&
  \begin{tabular}[t]{@{}l@{}}Weight discrimination accuracy was improved with active lifting vs.\\reflexive holding~\cite{brodie_sensorimotor_1984}.\end{tabular} \\ \hline
  
\multirow{5}{*}{\shortstack{Object \\ properties}} &
  Volume &
  \begin{tabular}[t]{@{}c@{}}Kinesthetic, tactile, \\ visual\end{tabular} &
  \begin{tabular}[t]{@{}l@{}}Smaller objects were perceived as heavier than larger objects~\cite{charpentier1886sensations}.\end{tabular} \\
 &
  Density &
  Visual &
  \begin{tabular}[t]{@{}l@{}}Denser looking objects were perceived heavier~\cite{wolfe_effects_1898}.\end{tabular} \\
 &
  Shape &
  \begin{tabular}[t]{@{}c@{}}Kinesthetic, tactile, \\ visual\end{tabular} &
  \begin{tabular}[t]{@{}l@{}}Objects with more compact shapes were perceived as heavier than \\less compact shapes with same weight and volume~\cite{dresslar1894studies}.\end{tabular} \\
 &
  Surface roughness &
  Tactile &
  \begin{tabular}[t]{@{}l@{}}Smoother objects were perceived as heavier than rough ones~\cite{flanagan_effects_1997}.\end{tabular} \\
 &
  Temperature &
  Tactile &
  \begin{tabular}[t]{@{}l@{}}Cold objects were perceived as heavier than warm ones~\cite{weber1905tastsinn}.\end{tabular} \\
\end{tabular}
\end{table*}

\subsection{Perception of weight \& contribution of tactile cues}
\label{sec:tactile_contribution}

When humans grasp, lift, and hold an object with their hand, the gravity acting on its mass creates a downward force (i.e., weight). For a successful lift or hold in the air, this weight should be stabilized with the friction force between the contacted skin and object, actively controlled by the grip force; see Fig.~\ref{fig:weight} for a schematic of these forces for a precision grasp. 

During lifting, humans perceive the object's weight by combining information from multiple sensory systems, predominantly somatosensory and visual ones~\cite{lim_systematic_2021}. The somatosensory system processes tactile cues, perceived through the skin receptors, and kinesthetic cues, sensed by the proprioceptors in muscle spindles or tendon organs. Furthermore, studies have shown evidence of the interplay between the two, where tactile mechanoreceptors responsible for skin stretch also contribute to kinesthetic cues by conveying information about joint angles~\cite{johansson_coding_2009, edin_quantitative_1992}. Even before lifting begins, individuals gauge an object's heaviness by scrutinizing its appearance. Upon touch and grasp, they acquire tactual information, such as contact shape, texture, temperature, and friction. During lifting, they perceive skin deformation, joint positions, and forces acting on muscles. Integrating all this sensory information by the central nervous system forms the basis for estimating an object's weight~\cite{jenmalm2000visual, johansson_coding_2009}. 

The role of each sensory information on perceived heaviness and how they are incorporated have been active research topics for nearly two centuries, dating back to the early psychophysical experiments conducted by Weber~\cite{weber_sense_1948}. He observed a significant discrepancy in weight discrimination between actively lifting objects by hand and passively perceiving them through cutaneous sensation when the hand was resting on a table. He found that active lifting was more than twice as precise, and this ability to discriminate masses through voluntary muscle exertion was called ``sense of force". Subsequent research highlighted the dominance of centrally generated motor commands in weight perception; a comprehensive review of these studies can be found in~\cite{jones_perception_1986}. 

While the proprioceptive sense has been consistently shown to play a crucial role in weight perception, various physical factors, e.g., biomechanical conditions and object properties, were also found to influence perceived heaviness~\cite{jones_perception_1986, lim_systematic_2021}. For a concise summary of these factors, the relevant sensory systems involved, and their impact on perceived heaviness, please refer to Table~\ref{tab:perception}. As indicated in this table, using different grasp conditions---e.g., wide vs. narrow---and lifting methods---e.g., active vs. reflexive holding---alter the perceived heaviness of objects, underscoring the important contribution of the tactile sense in weight perception. However, the integration mechanism of tactile and proprioceptive senses for making heaviness judgments is still an active research topic.

One of the early investigations into the contribution of tactile cues during a grasp-and-lift motion was conducted by Johansson and Westling~\cite{johansson_roles_1984}. The authors measured the grip forces of participants while manipulating small objects with different weights and surface frictions via pinch grasps. They showed that the participants' grip forces changed proportionally to load forces to overcome forces counteracting the intended manipulation. This balance between the grip and load forces was adapted based on friction to provide a small safety margin to prevent slips. They also demonstrated through experiments with local anesthesia that this adaptation occurred through cutaneous cues. 

In a subsequent study~\cite{westling_responses_1987}, they recorded afferent responses via microneurography when participants did the same grip-and-lift task, and they found activity in all four skin mechanoreceptors at different points. Fast adaptive (FA) I units were triggered during the object gripping and force oscillations during the holding phase. Slow adaptive (SA) I units responded during gripping and showed continuous firing during the static holding. As for the FA II units, they fired upon changes in contact or motion. Finally, the SA II units showed considerable sensitivity to skin deformation induced both by grip and load forces, indicated by an increased firing rate with force magnitude. These findings evidenced the contribution of skin receptors, particularly, SA II units, to the sense of weight. 

The contribution of tactile cues in weight perception was further proved by Jones and Piateski~\cite{jones_contribution_2006}. They conducted an experiment where participants produced forces with different muscle groups of the arm in the presence and absence of tactile stimuli and matched them using the corresponding muscle group in the other arm, always in the presence of tactile stimuli. Without tactile stimuli, participants tended to underestimate the reference forces for all muscle groups, with an increased effect of the perceptual detriment as weight increased. Similarly, in a recent study, Park et al.~\cite{park_effect_2020} observed that a rendered mass at the fingers was perceived to be heavier when tactile stimuli were present than in the condition with only kinesthetic feedback. Matsui et al.~\cite{matsui_relative_2014} attempted to measure the contribution ratios of tactile information and kinesthetic information to the perception of forces, obtaining a 16-28\,\% contribution of tactile information for a force of 1\,\si{N}, and 37--55\,\% for a force of 0.3\,\si{N}. However, this study only accounted for kinesthetic information from the finger and the hand since the rest of the arm was immobilized. The authors also noted that tactile and kinesthetic stimuli were only partially isolated. Recently, van Beek et al.~\cite{van_beek_static_2021} addressed the limitations of the study of Matsui et al. by utilizing a kinesthetic grounded robot to render the applied forces. Users pinch-grasped a manipulandum attached to the device's endpoint, under which force sensors were located. For isolating tactile and kinesthetic stimuli from each other, a pair of thimbles were worn to compress the finger and prevent tactile stimulation, while a padded finger rest was used to block kinesthetic information. The authors conducted multiple experiments to determine the DT, JND, and PSE under kinesthetic, tactile, and combined stimuli conditions upon presented weights. The authors reported that the combined condition yielded the lowest JND and DT, while the DT for the tactile condition was lower than the kinesthetic condition. Interestingly, while the provision of tactile information resulted in lower JNDs---i.e., provided more reliable information---compared to kinesthetic information for masses below 200\,\si{g}, both information sources were roughly equally reliable for larger weights. 

Interestingly, while the absence of tactile information decreases the ability to discriminate between small weights, the presence of tactile cues regarding object properties (see Table~\ref{tab:perception}) can also lead to misjudging object weight. This perceptual phenomenon, known as sensorimotor mismatch or \emph{weight illusions}, is still being investigated, and its underlying mechanisms are not fully understood. This phenomenon is generally associated with sensorimotor memory~\cite{buckingham2014illusions, saccone2019illusion}, which is captured by forward and inverse internal models, predicting the motor commands necessary to lift an object based on prior expectations of its weight and estimated uncertainty~\cite{Flanagan1997internal, Kawato2003forward, milner2006central}. Nonetheless, this perceptual deficiency has been utilized in practical applications for enhancing weight simulation within virtual environments through tactile interfaces~\cite{lim_systematic_2021}.

\section{Tactile weight rendering approaches}
\label{sec:tactile_solutions}
Overall, two leading weight rendering approaches through tactile stimulation have been proposed in the literature: inducing illusionary pulling sensations via \emph{asymmetric vibrations} and deforming the fingertips via \emph{skin stretch}. We further categorized the latter based on the mechanism that provided skin stretch, whether through the motion of \textit{flat surfaces} or \textit{belts} or \textit{tactors} actuated in planar/tangential or 3~DoF translational movements.

\subsection{Asymmetric vibrations}
\label{sec:vibrations}
\begin{figure}[b]
    \centering
    \includegraphics[width=9cm]{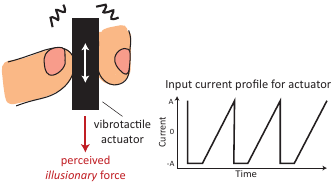}
    \caption{An illustration of weight rendering through asymmetric vibrations. A vibrotactile actuator moving on the vertical axis is held between the thumb and index finger. The actuator's input current, with negative values causing a downward acceleration, is designed to create an asymmetric acceleration pattern, stronger in a downward direction, leading to the perception of an illusionary downward force. The current profile is adapted from~\cite{culbertson_modeling_2016}.}
    \label{fig:asymmetric}
\end{figure}

This rendering approach relies on creating an illusionary pulling sensation that can modulate the perceived heaviness of an object. This illusion occurs by generating a higher peak acceleration or acceleration rate in one direction on the skin using a vibrotactile actuator, producing a sequence of alternating strong stimuli in one direction and weak stimuli in the other; see Fig.~\ref{fig:asymmetric}. The weak stimuli in this alternating pattern are not clearly perceived, resulting in a pulling sensation that increases the perceived weight of an object if the strong stimuli are in the gravity direction~\cite{tanabe_evaluation_2018, amemiya_asymmetric_2008}.

One of the first mentions of this effect dates back to Amemiya et al., who designed a slider-crank mechanism to induce a ``virtual force vector''~\cite{amemiya_virtual_2005}. This was achieved by oscillating the mass located at the endpoint of the mechanism with a peak acceleration along the major axis of the object much larger in one direction. Such an acceleration profile would, later on, be referred to as an \textit{asymmetric-in-amplitude} vibration, whose amplitude and frequency were shown to influence the intensity of the illusionary effect. In a posterior study, Amemiya and Maeda~\cite{amemiya_asymmetric_2008} demonstrated that their device could be used to perturb the perceived weight of an object when aligning the vibration in the direction of gravity. They observed that participants only perceived significant differences in weight when the pulling illusion was directed downwards (increased weight) but not upwards (decreased weight). The authors speculated that the strong additive force peaks resulting from the asymmetric acceleration profile were more saliently felt in the downward condition than the brief periods of reduced net force in the upward condition. 

Inspired by these works, Tappeiner et al.~\cite{tappeiner_good_2009} investigated how well this pulling illusion could be perceived in different directions in the horizontal plane. They used the Maglev device~\cite{berkelman1996designauthors}, a grounded magnetic levitation haptic interface, whose position was driven with a sinusoidal waveform with two half-periods of different durations to produce the asymmetric vibration. This acceleration profile was later referred to as an \textit{asymmetric-in-time} vibration for presenting a rate of change in acceleration more pronounced in one direction than in the other. The authors showed that users could guess the direction of vibrations with an error between 9 and 25 degrees.

To utilize the asymmetric vibration approach for weight rendering in mobile conditions, Rekimoto~\cite{rekimoto_traxion_2013} designed the Traxion device, which weighed only 5.2\,\si{g}. This development was a substantial improvement over previous solutions, which were much larger and heavier. This device utilized a linear electromagnetic vibration actuator (Force Reactor L-type, Alps Alpine, Japan) capable of inducing a pulling sensation with a much smaller size. By driving the actuator with a pulse-width modulated (PWM) signal, Rekimoto was able to achieve a pulling sensation of 0.292\,\si{N} with the device when held horizontally, perpendicular to gravity. Inspired by this innovation, researchers began exploring the use of alternative electromagnetic actuators, such as voice coil and linear resonance actuators, for implementing asymmetric vibrations.

Soon thereafter, Amemiya and Gomi~\cite{amemiya_distinct_2014} investigated the intensity of the pulling sensation elicited by asymmetric vibrations by employing the same actuator as~\cite{rekimoto_traxion_2013} and a voice coil actuator (Haptuator, Tactile Labs, Canada). Their findings revealed that the pulling illusion was most pronounced when the frequency of the asymmetric vibration signal coincided with the resonance frequency of the actuator. Another experiment within the same study showed that the illusion was most intense for the combination of the Haptuator and a driving signal of 40\,\si{Hz}. Follow-up work by Culbertson et al.\cite{culbertson_modeling_2016} carried out the dynamic modeling of this actuator and fingerpad to determine the optimal characteristics of the driving signal. Coinciding with the results from~\cite{amemiya_distinct_2014}, they identified an optimal input signal frequency of 40\,\si{Hz} for eliciting the pulling sensation. However, they proposed using a sawtooth step-ramp signal with a pulse width ratio of 0.3 to produce more asymmetry in the acceleration profile compared to a square wave signal (see Fig.~\ref{fig:asymmetric} for the proposed signal profile). These optimal configuration parameters were later used in two studies by~\cite{choi_grabity_2017} and~\cite{tanaka_dualvib_2020} on the use of asymmetric vibrations for weight rendering. 

One of those studies was performed by Choi et al.~\cite{choi_grabity_2017}, who designed Grabity, a haptic device for rendering grip contact and forces, weight, and inertia in a pinch grasp configuration. The device incorporated weight rendering capabilities through a pair of Haptuators aligned with the direction of gravity and driven by the signal proposed by~\cite{culbertson_modeling_2016}. The researchers demonstrated that the magnitude of the generated virtual forces can be adjusted by manipulating the amplitude of the asymmetric input signal. With their device, they could simulate increasing and decreasing virtual weight variations of up to 0.294\,\si{N} (30\,\si{g}). Notably, they observed that the perceived magnitude of the variations was smaller when decreasing the weight than when increasing it, in line with the results of~\cite{amemiya_asymmetric_2008}.
 
Later on, Tanaka et al.~\cite{tanaka_dualvib_2020} presented their DualVib device, a handheld device designed for rendering a \textit{dynamic mass} moving inside a container, such as a fluid or particles. The authors used asymmetric vibrations to render the forces of the moving mass using the actuator and driving signal from~\cite{culbertson_modeling_2016}, denoted as force feedback. In their design, they strategically positioned two Haptuators beneath the thumb and index fingers of the users to enhance the pulling sensation. Additionally, two electromagnetic vibration actuators (Haptic Reactors, Alps Alpine, Japan) were positioned on the palm for rendering the fine vibrations arising from the collisions of the dynamic mass with the container's inner surface, denoted as texture feedback. This distinction in actuation sites aimed to improve the perception quality of these two types of vibration stimuli. In their main experiment, participants were asked to distinguish between combinations of three different rendered materials (textures) through acoustic vibrations and three different mass levels rendered through asymmetric vibrations, both in isolation and combined. The participants were only capable of identifying conditions with an accuracy of $43.6\pm15.1\,\%$ for the combined one. A deeper look into the results showed that the combined condition yielded similar mass discrimination accuracy to the force-only condition. However, for material discrimination, the accuracy of the combined condition was closer to that of the texture-only condition. These results, together with the overall higher accuracy of the combined condition, suggest that asymmetric vibrations and texture feedback could be combined, or in terms of the authors, ``without mutual interference''~\cite{tanaka_dualvib_2020}. 

A common limitation pointed out in both aforementioned studies was the relatively small strength of the pulling sensation induced by the asymmetric vibrations, despite using the recommended combination of 40\,Hz signal and actuator of compatible frequency response by~\cite{culbertson_modeling_2016, amemiya_distinct_2014, choi_grabity_2017, tanaka_dualvib_2020}. More recently, Tanabe et al.~\cite{tanabe_evaluation_2018} provided further guidelines and observations about the usage of asymmetric vibrations, such as the minimum application time, the time until users develop perceptual adaptation to vibration, and further emphasis on designing the setup with a matching signal and actuator. In a follow-up study, Tanabe and colleagues created a voice-coil actuator of their own. They designed the output acceleration profile as a sinusoidal superimposed with its second harmonic with different phase shifts, resulting in various types of asymmetries~\cite{tanabe_pulling_2021}. The novelty introduced by the authors was the estimation of the combined fingertip and actuator transfer function per participant to ensure the accurate realization of the desired acceleration profile. By doing so, the authors observed that the pulling illusion was more strongly perceived when the resulting acceleration profile was asymmetric-in-time, with responses close to the chance rate for asymmetric-in-amplitude acceleration profiles. This result was further verified for different frequencies in a following study, which also confirmed that the best frequency for inducing the illusion was 40\,\si{Hz}~\cite{tanabe_effects_2020}.  

\begin{figure*}[t]
    \centering
    \includegraphics[width=\linewidth]{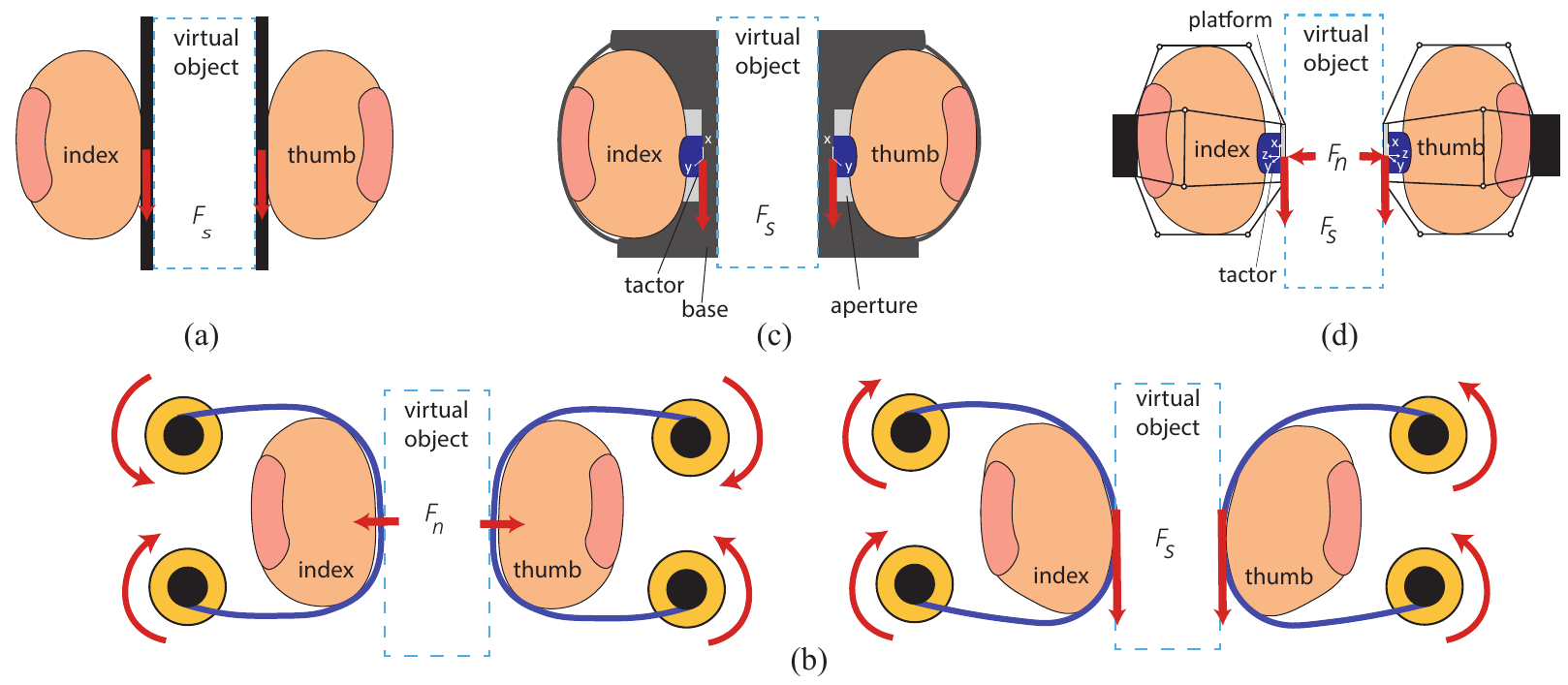}
    \caption{Illustration of weight rendering approaches through skin stretch. Here, the users grip virtual objects via precision grasp with their thumb and index fingers by wearing or holding the devices. (a) Skin stretch via \emph{flat surface motion}. e.g.,~\cite{kurita_weight_2011}. The object weight is simulated by controlling the displacement of the flat surfaces contacted with fingers, creating a shear force toward gravity. (b) Skin stretch through \emph{belt} motion, e.g.,~\cite{minamizawa_gravity_2007}. When the motors of a belt rotate in opposite directions and at the same rate, it deforms the corresponding finger only in the normal direction, simulating the normal stress due to grip. When they rotate in the same direction and rate, they deform the fingerpad only in the tangential direction, simulating the shear stress based on the desired weight of the virtual object. They can simultaneously deform the fingertip in normal and shear directions by rotating at different rates. (c) Skin stretch through a \emph{tactor} actuated in planar/tangential movement, e.g.,~\cite{girard_haptip_2016}. Each fingerpad rests on the base of the device and contacts the tactor through an aperture at the center. The displacement of the tactor creates a shear force, simulating the weight of an object. (d) Skin stretch via a \emph{tactor} actuated in 3~DoF translational movement, e.g.,~\cite{schorr_three-dimensional_2017}. The object weight is simulated by controlling the displacement of a tactor placed on a 3~DoF, kinematic delta structure, providing both shear and normal stress on the skin.}
    \label{fig:stretch}
\end{figure*}

\subsection{Skin stretch}
\label{sec:skin_stretch}

The second approach for rendering weight through tactile stimulation utilizes the deformations of the fingerpad caused by the tangential load forces during object lifting. This form of stimulation is widely known as skin stretch. Skin-stretch devices for weight rendering can be categorized based on the mechanism providing skin stretch, whether through the motion of \emph{flat surfaces} or \emph{belts} or \emph{tactors} actuated in planar/tangential or 3~DoF translational movements. Several studies employing these approaches for weight rendering are discussed in the following paragraphs. 

\subsubsection{Skin stretch through flat surface motion}
One mechanism to provide skin stretch for simulating weight in virtual environments is moving a flat surface in contact with the skin; see Fig.~\ref{fig:stretch}a for an illustration. Although such methodology has been used for rendering textures~\cite{wiertlewski2011encode} or understanding finger deformations~\cite{Tada2004image}, Kurita et al.~\cite{kurita_weight_2011} were the first ones to utilize it for weight rendering. Their one-DoF box-shaped device was held in a pinch grasp and worn through a pair of rings attached to the device. The transparent surface beneath the index fingertip was actuated vertically through a motor. During this motion, a camera captured the finger contact surface to calculate the fingerpad eccentricity, i.e., deformation. The virtual weights were rendered by controlling the position of the surface such that the measured fingertip deformation matched the average fingertip eccentricity profiles obtained by Mukai et al.~\cite{Mukai2009slip}, who measured the fingertip eccentricity of different participants while holding different weights. It should be noted that these average deformation profiles were used to simulate weights without accounting for differences in skin properties. The system was evaluated in an object identification experiment, in which participants grasped and lifted the device. On each trial, the device rendered the weight and friction coefficient of one of a set of real objects, which the user was asked to identify afterward. The authors observed that the device could render perceivably different levels of weight and friction, but the perceived values differed from the intended ones. Participants tended to rate the 100\,\si{g} object heavier than it was, while the 200\,\si{g} and 300\,\si{g} objects were underestimated. The authors discussed that such deviations could be attributed to the generalization of the deformation profiles, which did not account for individual differences in skin properties. 

\subsubsection{Skin stretch through belt motion}
\label{sec:belt}

Another way to render weight via skin stretch is by utilizing belt motion. One of the first skin stretch devices evaluated within this context was the Gravity Grabber, developed by Minamizawa et al.~\cite{minamizawa_gravity_2007, minamizawa_wearable_2007}. The device utilized two motors that actuated a fabric belt to deform the fingerpad skin in the vertical (normal) and shear (ulnar-radial) directions, as illustrated in Fig.~\ref{fig:stretch}b. When both motors rotated at the same rate in opposite directions, they induced vertical stress on the fingerpad, replicating the sensation of grasp contact and grip forces. Conversely, they produced shear stress when they rotated in the same direction and rate, simulating the weight perception. The authors showed that inducing various levels of shear stress on the fingerpad resulted in different perceived weights by the participants. The participants, who held the real objects in one hand, tuned the belt displacements for equivalent virtual weights (i.e., shear stress). From this experiment, they obtained a function that represented the relation between generated shear stress and the real object's weight. This function, an average gain factor for representing finger stiffness, was later used in follow-up experiments where they measured the reflexive response in grip force upon a sudden increase in real and simulated weights. The similar magnitude of the response in both conditions confirmed the suitability of the approach for weight rendering. 

Building upon their initial findings, Minamizawa et al.~\cite{minamizawa_simplified_2010} conducted experiments investigating the interaction between tactile and kinesthetic cues for weight rendering. They simulated object weight by displaying forces through a grounded kinesthetic device (Force Dimension, Omega 3), whose end effector was attached to the palm, wrist, or forearm via velcro straps. They placed urethane forms between the velcro straps and the skin to isolate the kinesthetic stimuli. They simultaneously deformed the fingerpad using their belt-based device placed on the fingertips. The first experiment aimed at measuring JNDs for reference stimuli between 50\,\si{g} and 400\,\si{g} using tactile cues alone or in combination with kinesthetic information applied at different locations, such as palm, wrist, and forearm. For stimuli below 200\,\si{g}, the tactile-only condition provided JND values comparable to those of the combined condition. However, for heavier stimuli, the JND increased largely for the tactile condition, indicating a stimulus saturation or a limitation in fingerpad deformation. The location of the kinesthetic stimuli did not affect the JND values. A similar result was observed in their second experiment, where the weight discrimination ability for a simulated object using tactile, kinesthetic (applied on the forearm), and combined cues was compared to that of a real object. First, the weight discrimination ability of the participants was measured while holding a cubic object attached to a grounded force feedback device (real object condition). JNDs were measured for reference stimuli of 100, 200, 300, and 400\,\si{g}. The experiment was then repeated while participants held the same object but the weight was simulated using tactile, kinesthetic, or combined cues. The JND values with the combined condition showed a similar trend to those measured with real objects, though the values were consistently slightly higher for all weights. The JND of the tactile condition was close to that of the combined condition for 100\,\si{g} objects, while the kinesthetic condition resulted in a JND almost twice as large. As the reference weight increased, the JND for the tactile condition increased substantially, while the kinesthetic condition approached that of the combined condition, matching it for the 400\,\si{g} reference stimulus. These results suggest that combining tactile and kinesthetic cues is particularly relevant for rendering lightweight objects. 

Besides the discussed implementation, other belt-based devices in the literature include the hRing, developed by Pacchierotti et al.~\cite{pacchierotti_hring_2016}. The device is worn on the proximal phalanges for compatibility with finger-tracking solutions. The authors demonstrated the device's capabilities for a pick-and-place task, obtaining overall lower interaction forces, completion time, and increased perceived effectiveness than in a visual-only condition. Other multi-cue devices featuring belt actuation have also been developed~\cite{murakami_altered_2017, vanriessen2023relocating}. However, these devices have yet to be evaluated within the context of weight rendering, particularly for those in which stimulation is relocated to the proximal phalanges. 

\subsubsection{Skin stretch through tactors actuated in planar/tangential movement}

The devices in this category generate skin deformations by displacing one or multiple small, high-friction tactile units (tactors) tangentially to the fingerpad. In these devices, the user's fingerpad rests on the contact base of the touchable, graspable, or wearable device~\cite{culbertson2018review}. Most designs involve an aperture on the base, allowing the fingerpad to directly contact the tactor(s), which are mechanically actuated beneath the base. The aperture also prevents the fingertip from moving in unwanted directions, thus ensuring that tactor displacements solely induce skin stretch; see Fig.~\ref{fig:stretch}c for an example of a wearable design. Unlike belt-based solutions, most tactor-based devices allow the rendering of tangential forces in 2~DoF, potentially covering the whole plane tangential to the fingerpad, which helps generate directional cues. The skin deformations due to the weight of objects can be simulated by controlling the displacement amplitude and speed of the tactor towards the direction of gravity (Fig.~\ref{fig:stretch}c). 

The earliest skin stretch devices that leveraged tactor actuation in planar/tangential movement were designed to provide directional cues on touchable device configuration~\cite{keyson_directional_1995, drewing_first_2005, gleeson_perception_2010}. The studies proposing these designs reported average direction discrimination thresholds between 11\si{\degree}~\cite{keyson_directional_1995} and 19\si{\degree}~\cite{drewing_first_2005} depending on the shear displacement amplitude and velocity, movement direction, and number of moving tactile units. Gleeson et al.~\cite{gleeson_perception_2010} showed that four orthogonal directions (distal, proximal, medial, and lateral) could be discriminated with high accuracy for displacements as small as 0.2--0.5\,\si{mm} and velocities of 1\,\si{mm/s} generated by one moving tactor. They also found that higher moving speeds and larger displacements caused greater accuracy in direction discrimination. Later, they showed the feasibility of using this rendering technique in a compact, fingertip-mounted design~\cite{gleeson_design_2010}. The results from these studies confirmed that moving small-sized tactile units could be exploited for rendering shear forces, like those occurring due to the weight of an object, in different directions on the fingerpad.  

Furthermore, Gleeson et al.~\cite{gleeson_improved_2011} proposed guidelines for designing skin stretch devices through tactor motion by testing tactors in two different textures and three sizes combined with apertures in three sizes in a direction discrimination study. They advised using rough-textured tactors to reduce slip and enhance direction identification accuracy. They found that the size of the tactor was insignificant for direction discrimination accuracy, suggesting that their size could be adaptable based on the application at hand and the finger size. Nevertheless, they recommended using a minimum tactor diameter of 7\,\si{mm} based on the reported user discomfort with smaller tactors. With regards to the aperture size, small sizes resulted in low direction discrimination accuracy, probably associated with the lack of stimulation of the skin surrounding the contact point. 

Following the design guidelines and results of Gleeson et al.~\cite{gleeson_perception_2010, gleeson_improved_2011}, further studies continued to explore the provision of directional cues with more compact devices~\cite{guinan_back--back_2013}. More prominently, some studies~\cite{schorr_sensory_2013, quek_augmentation_2014} aimed to render object stiffness using tactors actuated in tangential movement. These exemplary works used a graspable (i.e., stylus-like) skin stretch device with a tactor design following the guidelines from~\cite{gleeson_improved_2011}. They showed that their design could provide perceivable levels of stiffness comparable to grounded kinesthetic devices~\cite{schorr_sensory_2013} and augment the overall perceived stiffness when used in combination~\cite{quek_augmentation_2014}. These studies converted the desired rendering forces into tactor displacements by applying a predefined gain factor, similar to the belt-based devices described in the previous section. However, Quek et al.~\cite{quek_augmentation_2014} found that different gain factors resulted in different perceived stiffness levels with large intersubject variability. This variability was attributed to large differences in skin properties across participants, along with neural and cognitive factors. 

Despite their high shear force rendering capability, devices utilizing tactors actuated in planar/tangential movement have not been used for weight rendering applications until the design of the HapTip device~\cite{girard_haptip_2016}. HapTip features tactor actuation in the tangential plane of the fingertip in a wearable configuration (similar to the illustration in Fig.~\ref{fig:stretch}c), and it can render forces by projecting a weighted sum of gravity and inertial accelerations. The authors showed that HapTip could convey basic directions (up, right, down, and left) and orientations (horizontal, vertical, and two diagonals). They also evaluated the weight rendering capabilities of the device by embedding two HapTip devices to the two sides of a cube surface to provide tactile feedback (i.e., shear forces due to tactor displacement) on the thumb and index finger when the cube was held. In a virtual reality environment, they asked participants to sort virtual cubes based on their weight by picking and shaking each object. The weights of the virtual objects were arbitrary, had no real equivalents, and were adjusted based on the amplitude of the tactor displacements, such that their proportions were 1/3, 1, 3, and 9. The results showed that the most typical sorting error comprised the mistaking of the two heaviest cubes, attributed to the saturated actuation of the device. However, the overall positive results, with an average sorting error of 2.2 over 20, where 0 was the perfect ordering, supported the system's weight rendering suitability. 

\subsubsection{Skin stretch through tactors actuated in 3~DoF translational movement}\label{sec:parallel}

These devices create skin deformation through a tactor actuated via a 3~DoF (translational) kinematic mechanism, such as a delta structure~\cite{schorr_fingertip_2017}; see Fig.~\ref{fig:stretch}d for an illustration. This actuation allows the control of the tactor placed on a platform in all translational directions and, thus, can induce normal and tangential skin deformations. These devices have been designed in wearable or graspable configurations and broadly studied for various applications, such as rendering object curvature~\cite{prattichizzo_towards_2013}, stiffness~\cite{chinello_three_2018}, and weight. It should be noted that, despite their 3~DoF design, only the tactor motion towards gravity (shear force) is controlled for weight simulation. While this situation makes their rendering capability similar to the skin stretch through planar tactor motion, it also enables simulating contact and grasping forces when the user touches virtual objects. 

One of the earliest examples of this type of device used for rendering object properties was designed by Quek et al.~\cite{quek_sensory_2015}. The device was based on a delta mechanism; its end effector was connected to a rectangular tool having one tactor within one aperture on each of its four sides. The users grasped the tool with their thumb, index, and middle finger, hence each tactor stimulated a different finger. This configuration rendered forces on the fingerpads in both lateral and normal directions by moving the end effector, consequently tactors, in horizontal and vertical directions. They attached their tactile display to the end effector of a widely used kinesthetic haptic device (Force Dimension, Omega 3). Later, Suchoski et al.~\cite{suchoski_comparison_2016} used this device to render the mass of a virtual object when participants held the device by their thumb and index fingers in a pinch grasp. They compared the human perception of virtual masses rendered via their tactile display and the kinesthetic display. They measured the JND for the reference masses of 35, 70, 105, and 140~\si{g}. The participant's task was to adjust the mass of one block in a virtual environment in increments of 1~\si{g}, until it was equal to a reference block also being rendered. For the tactile display, the virtual masses were rendered by moving the tactors in the tangential direction based on a predefined gain factor (0.2\,\si{mm/N}) found by \cite{quek_sensory_2015}. They obtained Weber Fractions of 0.35 with the skin stretch device and 0.11 with the kinesthetic device, which the authors attributed to a more ``natural'' perception of the kinesthetic stimulation. It should be highlighted that although the device could render forces in 3~DoF translational movements, the weights were only rendered by 1~DoF (tangential) movement.

Recently, Schorr et al.~\cite{schorr_fingertip_2017, schorr_three-dimensional_2017} designed a compact, wearable device providing skin stretch through 3~DOF tactor motion (see Fig.~\ref{fig:stretch}d) and evaluated its weight rendering capability. They rendered virtual weights by converting the interaction forces in the virtual environment into tangential tactor displacements considering a predefined gain factor obtained from Nakazawa et al.~\cite{nakazawa2000shear}, and thus, regardless of the fingerpad stiffness variability among users. They conducted a weight magnitude estimation experiment using virtual objects of different sizes and weights. The results showed that participants could perceive differences in virtual object weight and that they applied increasing grasp forces when lifting virtual objects as rendered mass was increased. The same device was also used in a later study by Suchoski et al.~\cite{suchoski_scaling_2018} to evaluate the influence of scaling inertial forces on the perceived weight of a virtual object. The authors proposed using a scaling factor that multiplies the object's mass and divides the value of the gravitational acceleration in the virtual environment. By doing so, the overall weight of the object was preserved while the mass was increased, as did the inertial forces. They evaluated the effect of the scaling factor on weight perception in a discrimination experiment where participants were asked to pick and place two virtual objects: one reference object with 200\,\si{g} mass without scaling and one comparison object ranging between 50 and 350\,\si{g} with scaling factors of 2 and 3. They found that for a reference object of 200\,\si{g}, the average PSEs for the objects with scaling factors 2 and 3 were 171.0\,\si{g} and 150.5\,\si{g}, respectively. These results show that the proposed method could amplify the perception of the weight of virtual objects without increasing the force output of the device. 

The work of Leonardis et al.~\cite{leonardis_3-rsr_2017}, who used a parallel mechanism device with articulated legs connecting actuators to a platform, is relevant for its approach to dealing with varying finger properties. The authors performed per-participant fingerpad stiffness characterization with the device to ensure individualized conversion of the interaction forces into platform displacement, thus minimizing intersubject variability. They tested their device and approach for pick-and-place object manipulation. The results showed that participants reduced their grasping forces when using the device compared to visual information only, indicative of a better estimation of the object's weight. A significant difference in grip forces between light and heavy objects indicated the system's suitability for rendering different weight levels. Another experiment involving the lift-and-hold of an object restrained by a virtual prismatic constraint showed a similar pattern. The grasping and reaction forces generated by the prismatic constraint were significantly smaller when using the device than the visual condition.  

\section{Comparison of tactile weight rendering approaches}
\label{sec:comparison}

\subsection{Comparison criteria}
We compared the five different tactile weight rendering approaches---i.e., asymmetric vibrations and skin stretch through the motion of a flat surface or belt or tactors actuated in planar/tangential or 3~DoF translational movements---to guide the decision-making process for using them. To carry out the comparison, we established different criteria following the insights from the literature in the field and other properties listed in reviews on haptic devices and weight rendering, e.g., ~\cite{pacchierotti_wearable_2017, lim_systematic_2021, adilkhanov_haptic_2022}. 
 
The first two selected criteria are size and mass and relate to the physical properties of the devices used for tactile weight rendering. These characteristics condition the scope of usability of each approach; e.g., smaller systems are easier to accommodate, as well as wearables or hand-held devices. 

We also included criteria related to the mechanical characteristics of these devices, namely the number of actuated DoFs, maximum rendering force, and actuation workspace. The number of DoFs indicates the dimensionality in which a device can render weight and the possibility of simultaneous rendering of grip and load forces. For skin stretch devices, the maximum rendering force (i.e., the maximum weight the device can render) is also conditioned by the maximum skin stretch that a device can induce. Such a magnitude corresponds to its actuation workspace, which is additionally reported for this kind of device. 
 
The discrimination threshold for weight and directions constitutes another set of comparison criteria regarding the perceptual features of the devices. Lower discrimination thresholds and the capacity to render large rendering forces in a perceivable manner can translate into a device capable of providing a more naturalistic and coherent interaction. 

Finally, we added potential applications of the mentioned devices reported in the literature. 

\begin{table*}[t!]
    \centering
    \renewcommand{\arraystretch}{1.2}
    \caption{Comparison table of the five presented tactile weight rendering approaches across the selected criteria. Approaches for which a criterion does not apply have been categorized as ``Not applicable.'' Approaches for which no study has been found reporting a specific criterion have been labeled as ``Undetermined." Abbreviations: Degrees of Freedom (DoF),  Weight Discrimination Threshold (WDT), Direction Discrimination Threshold (DDT), Standard Deviation (SD). \dag Value converted to selected metric. \ddag Value with no conversion to chosen metric.}\label{tab:comparison_table}%
    \begin{tabular}{llllll}
        \hline
        \multicolumn{1}{c}{\textbf{Criterion}} 
        & \multicolumn{1}{c}{\begin{tabular}{@{}c@{}}\textbf{Asymmetric} \\ \textbf{vibrations}\end{tabular}} 
                         & \multicolumn{1}{c}{\begin{tabular}{@{}c@{}}\textbf{Flat surface} \\ \textbf{motion}\end{tabular}}
        & \multicolumn{1}{c}{\begin{tabular}{@{}c@{}}\textbf{Belt} \\ \textbf{motion}\end{tabular}} 
        & \multicolumn{1}{c}{\begin{tabular}{@{}c@{}}\textbf{Tactor motion} \\ \textbf{planar/tangential}\end{tabular}} 
        & \multicolumn{1}{c}{\begin{tabular}{@{}c@{}}\textbf{Tactor motion} \\ \textbf{3~DoF translational}\end{tabular}}
        \\

        \Xhline{3\arrayrulewidth}
        \multirow{2}{*}{\begin{tabular}{@{}l@{}}Size \\(\si{mm})\end{tabular}} & $7.5\times35.0\times5.0$                   ~\cite{rekimoto_traxion_2013}
                          & Undetermined
                  & $31.0\times28.0\times12.0$~\cite{pacchierotti_hring_2016}
                  & $20.4\times35.0\times34.1$~\cite{girard_haptip_2016} 
                  & $21.5\times48.8\times40.2$~\cite{schorr_fingertip_2017}
                  \\
                  & $10.0\times10.0\times35.0$~\cite{choi_grabity_2017} 
                  & 
                  &
                  & $45.9\times67.7\times18.5$~\cite{guinan_back--back_2013} 
                  & $18.0\times32.0\times32.0$~\cite{leonardis_3-rsr_2017}\\
    
                  & $56.0\times175.0\times27.0$~\cite{amemiya_asymmetric_2008}
                  & 
                  &
                  & $100.0\times32.0\times28.0$~\cite{quek_augmentation_2014}
                  & \\
    
        \hline
        \multirow{3}{*}{\begin{tabular}{@{}l@{}}Mass \\(\si{g})\end{tabular}} & $5.1$~\cite{rekimoto_traxion_2013} 
                                         &$210$~\cite{kurita_weight_2011}
                                & $15.3$~\cite{pacchierotti_hring_2016}  
                                & $22.43$~\cite{girard_haptip_2016} 
                                & $31.6$~\cite{schorr_fingertip_2017}
                               
\\
                                
                                & $8.15$~\cite{choi_grabity_2017} 
                                &  
                                &
                                &
                                & $16.31$~\cite{leonardis_3-rsr_2017} \\
                                &
                                &
                                &
                                &
                                & $260$~\cite{quek_sensory_2015}
                                \\
                                
        \hline
        \multirow{2}{*}{\parbox{0.1\textwidth}{DoF}} 
                        & $1$~DoF~\cite{choi_grabity_2017, tanaka_dualvib_2020, culbertson_modeling_2016}
                                                & $1$~DoF~\cite{kurita_weight_2011}
                        & $2$~DoF~\cite{minamizawa_gravity_2007, pacchierotti_hring_2016} 
                        & $2$~DoF~\cite{girard_haptip_2016, gleeson_design_2010}
                        & $3$~DoF~\cite{schorr_fingertip_2017, leonardis_3-rsr_2017} \\
                        & $2$~DoF~\cite{kim_hapcube_2018}
                        & 
                        &
                        &
                        & \cite{quek_sensory_2015, kamikawa_comparison_2018} \\
        \hline
        
        \multirow{4}{*}{\parbox{0.1\textwidth}{Maximum \newline rendering \newline force (\si{N})}} 
                        &  \dag $0.292$~\cite{rekimoto_traxion_2013}
                                                &Undetermined
                        &  \dag $3.92$~\cite{minamizawa_gravity_2007}
                        &  $3.4$~\cite{girard_haptip_2016}
                        &  $2.0\times2.0\times7.5$~\cite{schorr_fingertip_2017}
                        \\
    
                        &  \dag $0.294$~\cite{choi_grabity_2017}
                        &
                        &
                        &
                        &  $2.72\times2.73\times4.16$~\cite{leonardis_3-rsr_2017}\\
    
                        &  $0.43$~\cite{tanabe_evaluation_2018}
                        & 
                        &
                        &
                        &  $4.7$~\cite{chinello_design_2015} \\
                        &
                        &
                        &
                        &
                        &$5.0$~\cite{kamikawa_comparison_2018}
                        \\
                        
        \hline
        \multirow{4}{*}{\begin{tabular}{@{}l@{}}Workspace \\ (\si{mm})\end{tabular}} 
                        & Not applicable
                                               &Undetermined
                       & Unlimited
                       &  $\pm2.0$~\cite{girard_haptip_2016}
                       & $5.0\times10.0\times10.0$~\cite{schorr_fingertip_2017}
                       \\

                       & 
                       &
                       &
                       & $\pm2.5$~\cite{guinan_back--back_2013}
                       & $\pm7.5\times10.0\times15.0$~\cite{leonardis_3-rsr_2017} \\

                       &
                       &
                       &
                       & $\pm2.3$\cite{quek_augmentation_2014}
                       & $\pm\,\frac{\pi}{5}, \pm \frac{\pi}{6}, 15.0$~\cite{chinello_design_2015}  \\
                       &
                       &
                       &
                       &
                       & $5.0\times5.0\times5.0$~\cite{quek_sensory_2015}\\
    \hline
        \multirow{2}{*}{WDT}     
                        & Undetermined  
                                                &Undetermined
                        & {\begin{tabular}{@{}l@{}}\dag$0.1$--$0.25$~\cite{minamizawa_simplified_2010}\\ (ref. $50$--$300$\,\si{g})\end{tabular}} 

                        & Undetermined
                        & {\begin{tabular}{@{}l@{}}\dag$0.147$ \& $0.154$~\cite{suchoski_scaling_2018} \\ (ref.  $150.5$ \& $171.0$\,\si{g})\end{tabular}}
                        \\
                        &
                        &
                        &
                        &
                        & {\begin{tabular}{@{}l@{}}$0.35$~\cite{suchoski_comparison_2016} \\ (ref. $35$--$140$\,\si{g}) \end{tabular}}\\
                        
        \hline
        \multirow{3}{*}{\parbox{0.1\textwidth}{DDT}} 
                        & {\begin{tabular}{@{}l@{}}71\,\%\\ 8-direction \\ accuracy~\cite{kim_hapcube_2018}\end{tabular}}
                        & Not applicable
                        &Not applicable
                        &{\begin{tabular}{@{}l@{}}\dag84.7\% \\ 4-direction \\ accuracy~\cite{girard_haptip_2016}\end{tabular}}
                        & {\begin{tabular}{@{}l@{}}69\% \\8-direction\\ accuracy~\cite{leonardis_3-rsr_2017}\end{tabular}} 
                        \\

                        &  $^\ddag 9^{\circ}$--$25^{\circ}$ SD~\cite{tappeiner_good_2009}
                        &
                        &
                        & $12.9^{\circ}$--$15.6^{\circ}$\cite{keyson_directional_1995}
                        & \\
                        
                        & 
                        &
                        &
                        & $23^{\circ}$--$25^{\circ}$~\cite{drewing_first_2005}\\
                        \hline
        \multirow{3}{*}{\parbox{0.1\textwidth}{Potential \\ applications}} 
                        & {\begin{tabular}{@{}l@{}}Object exploration\\ in VR~\cite{amemiya_asymmetric_2008, choi_grabity_2017}\end{tabular}}
                        & {\begin{tabular}{@{}l@{}}Object exploration \\ in VR~\cite{kurita_weight_2011}\end{tabular}}
                        & {\begin{tabular}{@{}l@{}}Object exploration\\ and/or manipulation in \\VR/AR\cite{minamizawa_gravity_2007, pacchierotti_hring_2016, murakami_altered_2017, vanriessen2023relocating}\end{tabular}}
                        & {\begin{tabular}{@{}l@{}}Object exploration \\ in VR~\cite{girard_haptip_2016}\end{tabular}}
                        & {\begin{tabular}{@{}l@{}}Object manipulation \\and exploration in \\VR/AR~\cite{schorr_fingertip_2017, leonardis_3-rsr_2017, suchoski_scaling_2018}\end{tabular}}
                        \\
                        
                        & {\begin{tabular}{@{}l@{}}Direction communication \\~\cite{amemiya_virtual_2005, tappeiner_good_2009}\end{tabular}}
                        &
                        &
                        &\begin{tabular}{@{}l@{}}Sensory substitiuon \\ in teleoperation\cite{schorr_sensory_2013, quek_augmentation_2014}\end{tabular}
                        &{\begin{tabular}{@{}l@{}}Sensory substitiuon \\ in teleoperation~\cite{quek_sensory_2015}\end{tabular}}
                        \\
                        
                        &
                        &
                        &
                        & {\begin{tabular}{@{}l@{}}Direction communication \\~\cite{keyson_directional_1995, drewing_first_2005, gleeson_design_2010}\end{tabular}}
                        
                        &
    \end{tabular}
\end{table*}

\subsection{Comparison}
Table~\ref{tab:comparison_table} summarizes our comparison results of tactile weight rendering approaches. Here, we report the results using a standard metric for each listed criterion, converting other reported metrics when possible. For the size and the mass, we report magnitudes as indicated by the authors of the device, typically as the length, width, and depth in millimeters, and the mass in grams. For the number of DoFs, we considered the values reported by the authors and otherwise derived them from the actuation of the system and the kinematic model. 

We computed the maximum rendering forces for asymmetric vibration devices as the maximum perceived force of the pulling illusion. Whenever the maximum rendering was reported by the authors as the rendered mass, we converted the value to force unit, Newton. 

The workspace, just like the size, is typically listed as the range in each actuated dimension. However, we did not evaluate the workspace of asymmetric vibration devices, since the actuators vibrate in place and transmit the vibrations to the device they are attached. Thus, their effective workspace is negligible. For belt-based devices, on the other hand, we indicated that they provide an ``unlimited'' workspace in the actuated DoF(s). In the normal and ulnar-radial directions, the workspace limits of the device depend on the length of the actuated fabric belt and constraints of the finger deformation. Therefore, provided sufficient force by the actuators, the device can potentially deform skin up until its maximum deformation, which some articles have measured to be up to 5\,\si{mm} in the shear direction~\cite{pataky_viscoelastic_2005}, or until slippage occurs. 

We present the weight discrimination threshold (WDT) as the Weber Fraction (WF) with the corresponding reference weights. Whenever the value was provided as the JND, we computed the WDT by dividing the JND by the reference weight. We did not provide values for asymmetric vibrations and skin stretch through flat surface motion as those studies looking into weight rendering typically do not investigate discrimination thresholds but compare two or three weight levels or shifts in perceived weight~\cite{choi_grabity_2017, tanaka_dualvib_2020, kurita_weight_2011}. Although these results indicate the rendering capability of the device, none of the reviewed studies provides a clear metric for the weight discrimination threshold, so we considered the value to be unknown.

For the direction discrimination threshold (DDT), however, we did not consider belt and surface devices for this criterion since they are only capable of rendering tangential forces in a single direction. We provided two types of metrics for the remaining approaches: the direction detection accuracy and the JND, depending on the experimental approach. One result worth elaborating on is the value provided for asymmetric vibrations, obtained from the study of~\cite{tappeiner_good_2009}. In the experiment, participants were presented with a vibration and were allowed to freely indicate its direction. The reported values in the table correspond to the mean of the within-subject standard deviation when indicating the direction of the vibrations.

For the potential applications, we consider the target application of the studies developing and evaluating the devices. Additionally, we also considered the potential applications discussed by the corresponding authors for proof-of-concept and technical studies. 

\section{Discussion}
\label{sec:discussion}
This review addresses the usage of tactile interfaces for rendering the weight of virtual objects. These devices can substitute and augment kinesthetic devices to provide a more coherent, accurate, and naturalistic sensation of weight. Doing so can potentially increase manipulation performance in teleoperation tasks or improve the efficacy of robot-assisted and VR simulators for training and learning motor tasks. We first classified the tactile weight rendering approaches found in the literature within a proposed categorization. Then, we compared them across several criteria based on the insights of the retrieved studies to provide information to allow the selection of approaches based on developers' requirements. 

\subsection{Which approaches have been used to render weight through tactile stimulation?}

Two major groups of approaches have been determined according to the type of stimulus used to render weight: asymmetric vibrations and skin stretch. The asymmetric vibration approach can simulate weight by inducing a pulling sensation by actuating a vibration motor with an asymmetric output acceleration profile. The skin stretch approach is based on deforming the fingerpad to stimulate the skin mechanoreceptors and induce the sensation of weight. This approach can rely on the actuation of a flat surface, belt, or tactor in planar/tangential or 3~DoF translational movements. Multiple studies have been presented evaluating these approaches in grounded, hand-held, and wearable devices in the context of weight rendering.

It should be noted that different categorizations of these approaches can also be performed. For example, Pacchierotti et al. classified haptic devices according to whether they were worn on the hand or the fingertip~\cite{pacchierotti_wearable_2017}. Hand devices were further separated by kinesthetic or vibrotactile feedback, while fingertip devices were divided according to the rendered sensation (e.g. normal indentation, lateral skin stretch). Lim et al. classified weight rendering devices according to the haptic cue used by the devices, namely forces, skin stretch, vibrations, weight shifting, and others~\cite{lim_systematic_2021}. Finally, Adilkhanov et al. classified the devices in their review based on the degree of wearability, further classified by their actuation principle~\cite{adilkhanov_haptic_2022}.

The approaches presented in this review are classified similarly to the review by~\cite{lim_systematic_2021}, limited to vibrations and skin stretch, which act upon the tactile sense. However, from the whole group of vibration approaches, the focus has been directed toward asymmetric vibration systems since they are the only vibration solution that renders a perceivable force or weight sensation rather than a cue proportional to the object's weight. The latter approach, although indicating the weight of an object, does not actually render weight and, thus, was excluded from this review. The skin stretch approach, following a similar perspective to~\cite{adilkhanov_haptic_2022}, has been further divided according to the actuation or stimulation mechanism of the available devices. Such a perspective was followed instead of a sensation-based classification, as in~\cite{pacchierotti_wearable_2017}, because most of these devices can provide a combination of sensations, such as normal indentation and lateral skin stretch. Additionally, although some of the devices are inherently wearable, like the ones that rely on belt motion, others can be integrated into all sorts of solutions, like tactor devices, implemented in wearable~\cite{girard_haptip_2016} and grounded solutions~\cite{quek_sensory_2015}. Classification according to the actuation principle allows for a comparison of devices that can be used across all wearability scales, increasing the scope of application of the results.

\subsection{What are the main advantages and disadvantages of each approach? }

The comparison performed in the previous section showed that each approach presented several advantages and limitations, indicative of better suitability of each approach to specific applications or contexts. Ideally, an experimental comparison of the available devices is needed to determine the best approach to each case using different experiments related to the application of interest. However, as far as we know, no such comparative studies for weight-rendering exist in the literature. 

Despite the absence of comparative studies providing a quantitative and thorough comparison of approaches per application, our results indicate the applications in which each approach can perform best. Asymmetric vibration devices with small sizes and weights come up as a solution mainly developed and suited for handheld devices, for being more straightforward to integrate into those systems, such as VR controllers. However, factors like the limited rendered weight and the perceivable vibratory effect from the actuators limit their versatility~\cite{choi_grabity_2017, tanaka_dualvib_2020}. Therefore, we anticipate that recent skin stretch devices with small form factors, like the Chasm device~\cite{preechayasomboon_chasm_2020}, can potentially replace asymmetric vibration devices due to their larger actuation forces and small size.

Only one study has been found exploring the induction of skin stretch through flat surface motion~\cite{kurita_weight_2011} for weight rendering. This approach can deform the entire fingerpad in the movement direction, thus potentially inducing a more naturalistic sensation. Due to its actuation principle, it can potentially generate 2~DoF tangential deformations, offering a more versatile solution. The main limitation of this approach, and a possible explanation for the reduced number of studies exploring it, is the difficulty of immobilizing the finger to induce skin stretch upon movement of the surface exclusively. Grounding the device at more proximal phalanxes~\cite{minamizawa_gravity_2007, schorr_three-dimensional_2017} or integrating it into hand-worn devices~\cite{pacchierotti_wearable_2017} can increase the usability of this approach and the exploration of its weight rendering capabilities.

Among skin stretch devices, the ones utilizing belt motion present several advantages indicating their suitability for rendering object weight in virtual environments. Belt devices constitute a highly and strictly wearable solution that has shown good performance at simulating the weight of virtual objects. In addition to displaying the largest rendered weight observed across all studies in this review, these devices yield comparable weight discrimination thresholds in the presence and absence of kinesthetic information for low weights~\cite{minamizawa_simplified_2010}. Their main limitation is the single degree of actuation in the ulnar-radial direction, which prevents the rendering of weight and inertial forces in the proximal-distal direction of the fingerpad. Hence, they restrict the grasping configuration for lifting the object and could cause a mismatch of sensory information during object manipulation. Overall, the wearable and compact form factor of belt devices makes them better suited for augmented and virtual reality applications that require free hand movements and relatively large weights to be rendered. However, the grasping motion of the object needs to be restrained to a lateral grasp for the feedback to be effective.

The use of tactors actuated in planar/tangential motion can compensate for the lack of stimulation in the proximal-distal direction of the fingerpad. As a result, cues (like weight) can be conveyed in multiple directions~\cite{tappeiner_good_2009, drewing_first_2005, keyson_directional_1995}, thus supporting a wider range of lifting configurations. The main limitation of this approach resides in the grounding of the outer part of the fingerpad, which can lead to reduced perceptual acuity, as discussed in the incoming paragraphs. This grounding, on the other hand, allows the rendering of feedback without attaching the finger to the device. For this reason, these devices are most studied for teleoperation~\cite{quek_augmentation_2014}.

The wearable skin stretch devices relying on tactor actuation in 3~DoF translational movements provide an additional level of rendering capabilities. The three-DoF actuation of these devices allows the rendering of forces in three translational dimensions, enabling the rendering of weight and normal forces and simulating contact and non-contact situations in all grasp configurations. With their reduced discrimination thresholds, these devices present an excellent all-around performance. All this comes at the cost of reduced tangential forces and a bulkier device compared to the devices relying on tactor actuation in planar/tangential movement. Therefore, the available wearable skin stretch devices utilizing tactor motion in 3~DoF allow the rendering of weight in virtual reality applications.

The skin stretch device by Suchoski et al., utilizing tactor actuation in 3~DoF translational movements with graspable configuration, has shown the most limited weight discrimination thresholds among skin stretch devices~\cite{suchoski_comparison_2016}. Although the maximum rendering force achieved with this device is large, the maximum rendered weight found in the literature is usually much smaller. The reason for both of these observations, as discussed in~\cite{suchoski_comparison_2016}, can be attributed to the local deformation of the skin in contact with the tactor relative to the surrounding skin, grounded on the aperture. Because a large part of the fingerpad is grounded in the system's aperture, the corresponding mechanoreceptors are not stimulated, potentially contributing to the reduced discrimination accuracy and maximum rendered weight. This observation is backed up by~\cite{gleeson_improved_2011}, who noticed a reduction in direction discrimination performance for smaller apertures, which constrained a larger area of the skin.

In summary, each weight rendering approach offers distinct advantages alongside corresponding disadvantages, requiring careful consideration tailored to specific use cases and limitations. 

\subsection{Limitations of current approaches and future perspectives}
Below, we propose future research directions on tactile weight rendering based on the limitations of current approaches found in the literature. Further research in these directions could ultimately increase the performance of these tactile weight rendering approaches, impacting user experience, immersion, and performance during training and teleoperation tasks. 

\subsubsection{There is a need for comparative studies and standardization} As mentioned earlier, providing a definitive best tactile weight rendering approach for specific applications is challenging due to the need for more comprehensive information on particular properties of the reported devices in the literature and comparative studies that experimentally evaluate and compare the existing devices. As it may not always be feasible to access these devices, one way to mitigate this problem is by standardizing the process, such as developing design and evaluation guidelines, open-source software and hardware for testing, and information-sharing platforms, with a joint effort of experts.

\subsubsection{Devices should account for variability in fingertip properties} In the studies conducted with skin stretch devices, e.g.,~\cite{quek_augmentation_2014, kurita_weight_2011}, large intersubject variability was observed in the perceived properties of the virtual objects and task performance. One reason for this variability was the use of position control, which relied on a constant scaling factor or an average fingerpad stiffness to convert the virtual forces into displacements, disregarding differences in skin properties across participants. Several studies reported that such differences in finger mechanical properties and size could cause large variabilities in fingertip deformation~\cite{serhat2022contact} and resulting tactile perception~\cite{richardson2022learntofeel, nam2020stickiness}. Therefore, these variabilities in skin properties should be considered for interface design. For skin-stretch devices, one idea to overcome this problem could be the development of force-controlled systems, as done by~\cite{kamikawa_comparison_2018, ratschat2024design}, or per-participant estimations, as in the study of~\cite{leonardis_3-rsr_2017}. For asymmetric vibrations, such variabilities could be mitigated by designing the acceleration signal considering the differences in fingertip properties in the dynamic model of finger-actuator contact, as proposed by~\cite{tanabe_pulling_2021}. 

\subsubsection{Devices should allow more diverse grasp configurations} Most of the reported devices in this review allow users to interact with virtual objects in a pinch grasp, with the index finger and the thumb in opposition. As it is known that the number of fingers used to lift an object influences the perceived weight~\cite{flanagan_coming_2000}, more research is needed on developing devices supporting power and multi-finger grasp and evaluating existing solutions in different exploration and lifting strategies. 

\subsubsection{There is a need for realistic rendering of object properties} Most studies in this review evaluated tactile interfaces in terms of their capability to simulate perceivable weights---not necessarily counterparts of realistic objects. A closer resemblance between virtual and physical objects can increase immersion~\cite{newman_use_2022} or a better transference of learning in a virtual environment to the real world~\cite{levac2019learning}. Hence, realistically rendering object properties, such as exact weight, friction, or texture, is another exciting future research direction. However, such a multisensory aim requires a technological leap in device design and control. 

\section{Conclusions}
\label{sec:conclusion}

This study provides an overview and comparison of weight rendering approaches through tactile stimulation as a potential approach to enhance rendering accuracy when used in conjunction with kinesthetic devices or to substitute kinesthetic stimulation when simulating small weights. We conducted an exhaustive literature search and proposed a categorization of the different approaches followed by a comparison across several criteria based on the insights of the retrieved studies. This search distinguished two main approaches: asymmetric vibrations and skin stretch, induced via the motion of a belt or flat surfaces or tactors actuated in planar/tangential or 3~DoF translational movements. Based on the comparison, the asymmetric vibration approach provides some limitations that indicate that its use for weight rendering is limited to applications involving tight size constraints and low-fidelity rendering. Although each skin stretch device has specific advantages, the large maximum rendering force and low weight discrimination threshold, among others, indicate increased suitability of belt and 3D tactor devices for weight rendering. The limitations of the solutions and gaps in the literature identified in this review indicate a need for further research to determine the optimal way of rendering weight via the tactile sense. This study aims to motivate and guide the development and usage of tactile displays for more accurate weight rendering, improving immersion in virtual reality and performance and learning in training and teleoperation applications.

\bibliographystyle{IEEEtran}
\bibliography{references}

\begin{IEEEbiography}
[{\includegraphics[width=1in,height=1.25in,clip,keepaspectratio]{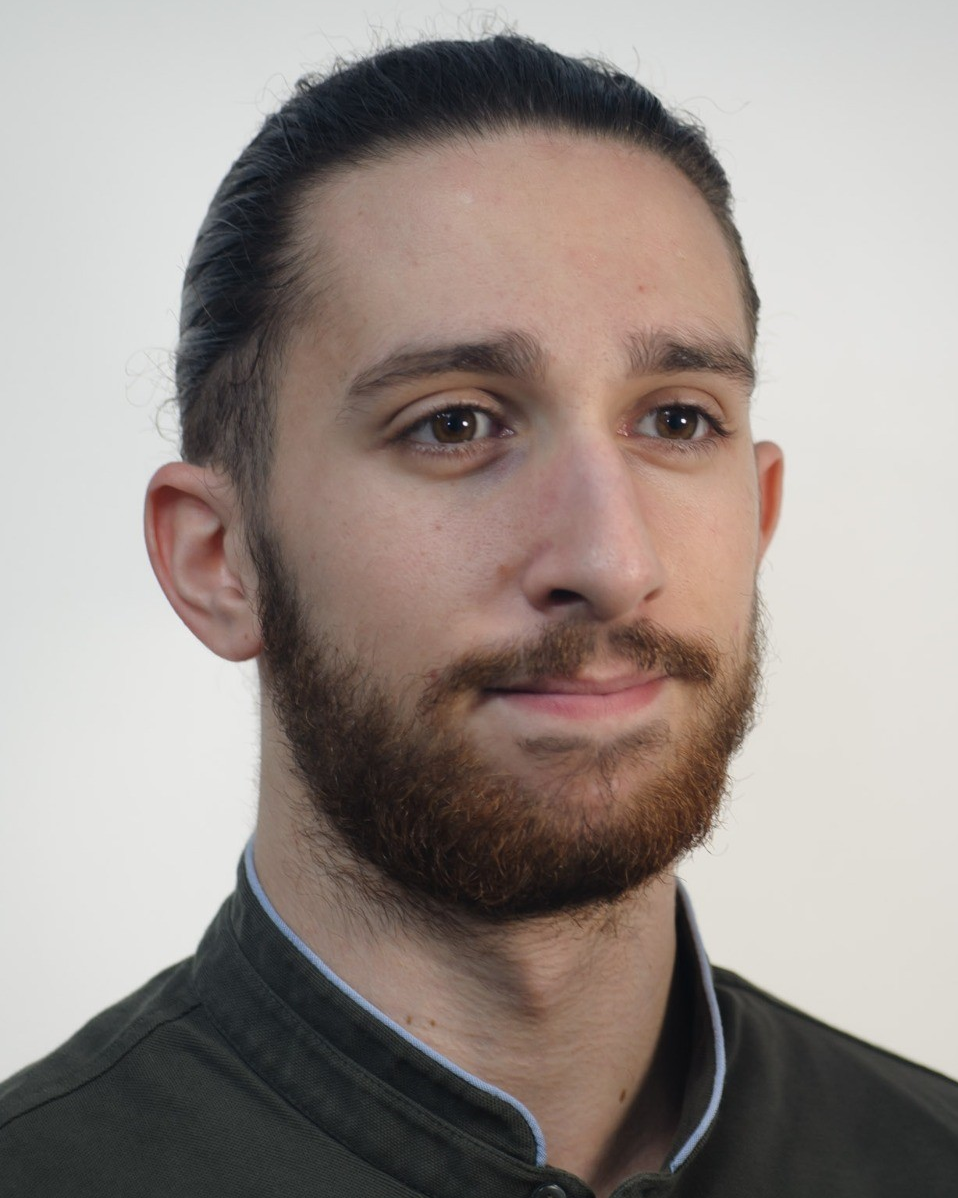}}]{Ruben Martin Rodriguez}
is a PhD candidate at the University of Basel (Switzerland). Previously, he earned his MSc in robotics at Delft University of Technology (The Netherlands) and his BSc in industrial electronics and automation engineering at Universidad Carlos III de Madrid (Spain). His early research focuses on the development of haptic systems for rehabilitation. Currently, his research interest focuses on the development of robotic solutions for in situ 3D bioprinting. 
\end{IEEEbiography}

\begin{IEEEbiography}
[{\includegraphics[width=1in,height=1.25in,clip,keepaspectratio]{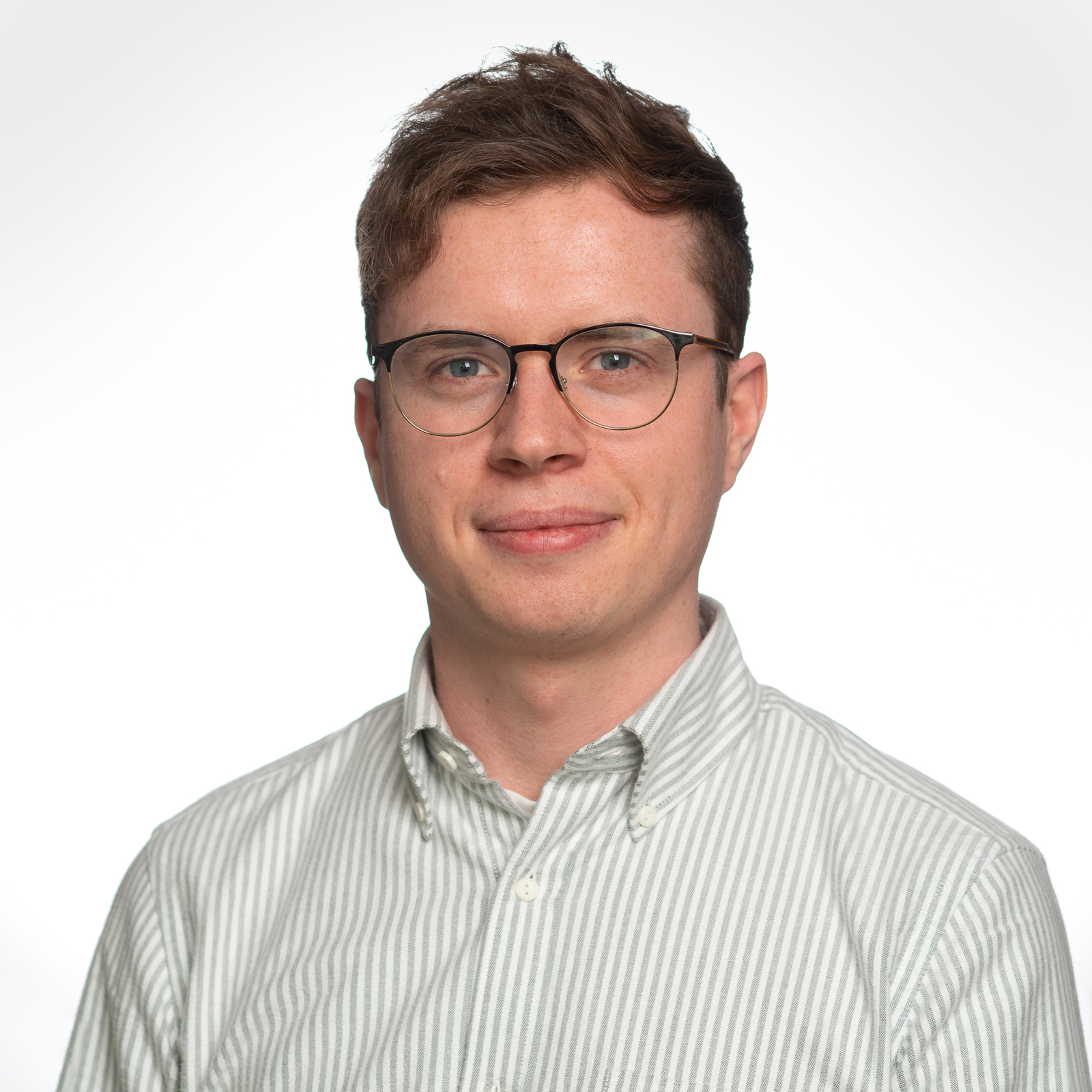}}]{Alexandre L. Ratschat}
is a Ph.D. candidate at the Delft University of Technology (The Netherlands). He is also affiliated with the Department of Rehabilitation Medicine, Erasmus Medical Center (The Netherlands). He obtained his B.Sc. and M.Sc. in mechanical engineering from ETH Zurich (Switzerland). From 2019--2020 he worked at F\&P Robotics AG (Switzerland) as a software engineer. He started his position as a Ph.D. candidate in February 2022, where his research focuses on leveraging robotic devices for upper-limb rehabilitation and motor learning of people with acquired brain injury. 
\end{IEEEbiography}

\begin{IEEEbiography}
[{\includegraphics[width=1in,height=1.25in,clip,keepaspectratio]{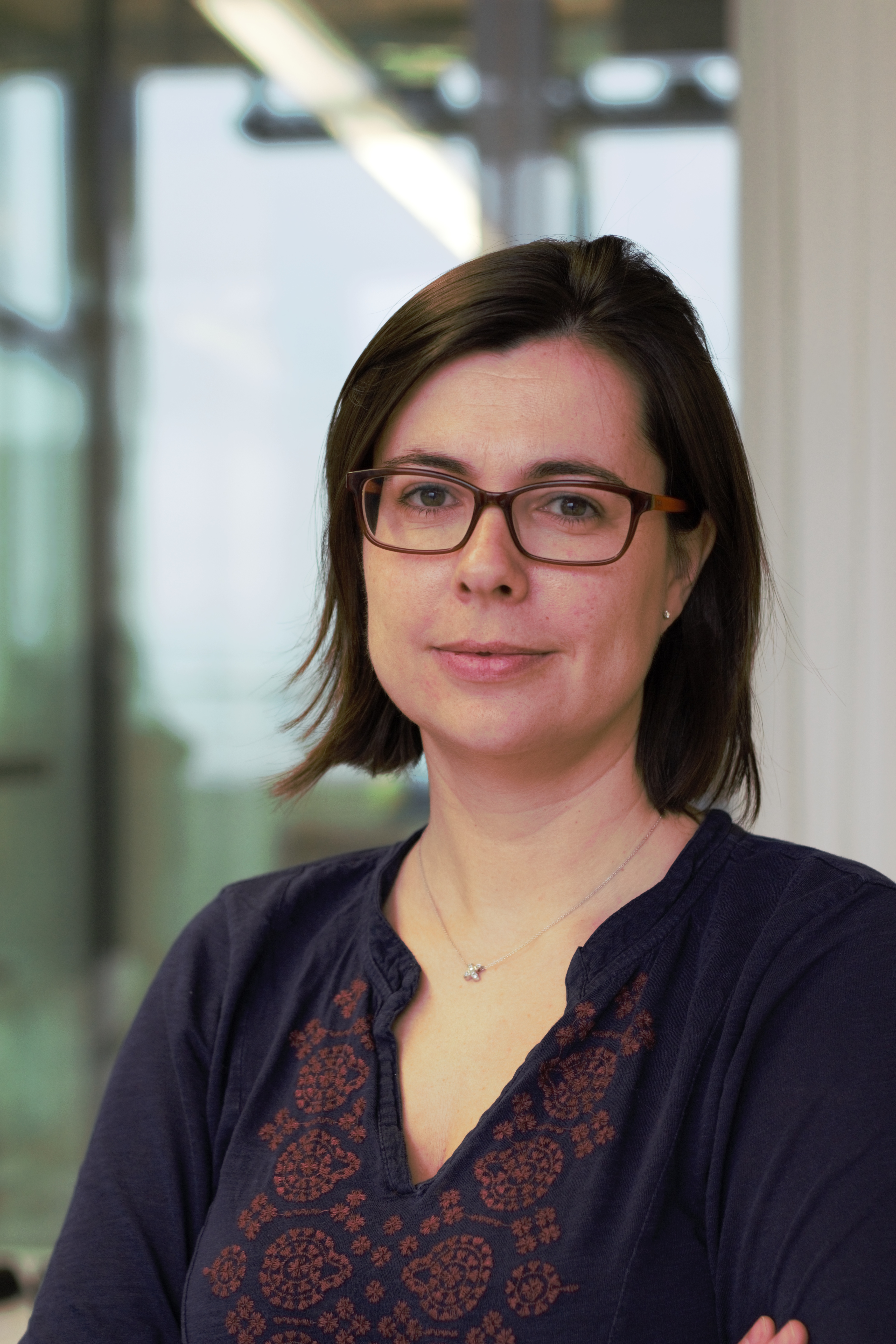}}]{Laura Marchal-Crespo}
is an Associate Professor at the Delft University of Technology (The Netherlands). She is also affiliated with the Department of Rehabilitation Medicine, Erasmus Medical Center (The Netherlands), and the University of Bern (Switzerland). She obtained her B.S. in industrial engineering from the Universitat Politecnica de Catalunya (Spain) and M.Sc. and Ph.D. degrees from the University of California at Irvine (USA). She then joined ETH Zurich (Switzerland), as a postdoc researcher. In 2017 she obtained a Swiss National Science Foundation (SNSF) Professorship and joined the University of Bern as medical faculty. She became an Associate Professor at the Delft University of Technology in September 2020. She carries out research in the general areas of human-machine interfaces and biological learning, and, specifically, in the use of robotic assistance and virtual reality to aid people in learning motor tasks and rehabilitate after neurologic injuries.
\end{IEEEbiography}

\begin{IEEEbiography}
[{\includegraphics[width=1in,height=1.25in,clip,keepaspectratio]{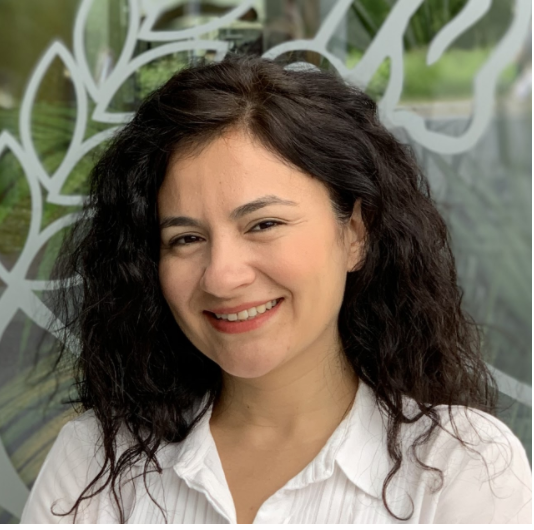}}]{Yasemin Vardar}
is an Assistant Professor at the Delft University of Technology (The Netherlands). Previously, she was a postdoctoral researcher at the Max Planck Institute for Intelligent Systems (Germany). She earned her Ph.D. in mechanical engineering at Koç University (Turkey), MSc. in systems and control at Eindhoven University of Technology (The Netherlands), and BSc. in mechatronics engineering at Sabanci University (Turkey). Her research interests focus on understanding human touch and haptic interface technology development. She received several awards, including the NWO VENI Award (2021) and Eurohaptics Best Ph.D. Thesis Award (2018). She is currently a co-chair of the Technical Committee on Haptics. 
\end{IEEEbiography}

\end{document}